\documentclass{article}
\usepackage{graphicx}
\usepackage{amsfonts}
\usepackage{amsmath}
\usepackage{amssymb}
\begin{document}

\title{Using Relative Entropy to Find Optimal Approximations:\\
an Application to Simple Fluids\thanks{Accepted for publication 
in Physica A, 2008}}

\author{Chih-Yuan Tseng\thanks{He is cuurently in the transition to Department of Oncology, University of Alberta, Edmonton AB T6G 1Z2; E-mail: richard617@gmail.com}\\
Graduate Institute of Systems Biology and Bioinformatics,\\ National Central University, Jhongli 320, Taiwan\\
\and Ariel Caticha\thanks{E-mail: ariel@albany.edu}\\
Department of Physics, University at Albany-SUNY\\ Albany, NY 12222 USA}
\date{}
\maketitle
\begin{abstract}
We develop a maximum relative entropy formalism to generate optimal
approximations to probability distributions. The central results consist in
(a) justifying the use of relative entropy as the uniquely natural criterion
to select a preferred approximation from
within a family of trial parameterized distributions, and (b) to obtain 
the optimal approximation by marginalizing over parameters using the method of 
maximum entropy and information geometry. 
As an illustration we apply our method to simple fluids. The
``exact'' canonical distribution is
approximated by that of a fluid of hard spheres. The proposed method first 
determines the preferred value of the hard-sphere diameter, and then obtains 
an optimal hard-sphere approximation by a suitably weighed
average over different hard-sphere diameters. This leads to a considerable
improvement in accounting for the soft-core nature of the interatomic
potential. As a numerical demonstration, the radial distribution function and the
equation of state for a Lennard-Jones fluid (argon) are compared with
results from molecular dynamics simulations.\newline
\end{abstract}



\section{Introduction}

\label{intro} A common problem in statistical physics is that the
probability distribution functions (PDFs) are always too complicated for practical calculations
and we need to replace them by more tractable approximations. A possible
solution is to identify a family of trial distributions $\{p(x)\}$, where 
 x are parameters characterized systems and
select the member of the family that is closest\ to the exact\ distribution $%
P(x)$. The problem, of course, is that it is not clear what one means by
`closest'. One could minimize

\begin{equation}
\int dx\,\left[ p(x)-P(x)\right] ^{2},
\end{equation}%
but why this particular functional and not another? And also, why limit
oneself to an approximation by a single member of the trial family? Why not
consider a linear combination of the trial distributions, some kind of
average over the trial family? But then, how should we choose the optimal
weight assigned to each $p(x)$? We propose to tackle these questions using
the method of Maximum relative Entropy (which we abbreviate as ME) and
information geometry \cite{Cencov81}. The ME method, 
which is developed in  \cite{ShoreJohnson80}-\cite{Caticha03b}, has
historical roots in the earlier method of maximum entropy that was pioneered
by E. T. Jaynes and is commonly known as MaxEnt  \cite{Jaynes57}. 
The ME method is designed for updating probabilities from arbitrary priors for information in
the form of arbitrary constraints and it includes Bayes' rule and the older
MaxEnt as special cases \cite{Caticha03}, \cite{Caticha03b}.



The purpose of this paper is to develop a ME based method to 
generate optimal approximations (brief accounts of some of the results discussed below
have previously been presented in \cite{Tseng02} and \cite{Tseng03}). 
The general formalism, which is the main result of this paper, is developed
in section \ref{sec2}. In section ~\ref{sec2-1} we justify the use of
relative entropy as the unique and natural criterion to select the preferred
approximation, which is labeled by some parameters. 
The optimal approximation is
obtained in section \ref{sec2-3} by marginalizing over the variational
parameters. A suitably weighed average over the whole family of trial
distributions with the optimal weight provides an optimal approximation.

In the second part of the paper we demonstrate the proposed formalism 
by applying it to simple classical
fluids, a well studied field in the past \cite%
{BarkerHenderson76}-\cite{Kalikmanov02}.  
To approximate the behavior of simple fluids we chose trial distributions
that describe hard spheres \cite{BarkerHenderson76}-\cite{Kalikmanov02}
(section ~\ref{sec3}). The ME formalism is first used (section ~\ref{sec4}) to
select the preferred value of the hard-sphere diameter. This is equivalent to
applying the Bogoliubov variational principle and reproduces the results
obtained by Mansoori et al. \cite{Mansoori69} whose variational principle
was justified by a very different argument.

An advantage of the variational or the ME methods over the perturbative
approaches such as Barker and Henderson (BH) \cite{BarkerHenderson76} and
of Weeks, Chandler and Anderson (WCA) \cite{WCA} is that there is no 
need for ad hoc criteria dictating how to
separate the intermolecular potential into a strong short range repulsion
and a weak long range attraction. On the other hand, a disadvantage of the
standard variational approach is that it fails to take the softness of the
repulsive core into account. At high temperatures this leads to results that
are inferior to the perturbative approaches. 

In the standard variational a single preferred value of the hard-sphere
diameter is selected. But, as discussed in \cite{Caticha03} and \cite%
{Caticha00}, in the ME method non-preferred values are not completely ruled
out. This allows us (in section ~\ref{sec5}) to marginalize over hard sphere diameter 
to obtain an optimal hard-sphere approximation with a suitable weighting. 
That leads to significant improvements over
the standard variational method.

In section ~\ref{sec6} we test our method by comparing its predictions for a
Lennard-Jones model for argon with molecular dynamics simulation data (\cite%
{Verlet68}, \cite{HansenVerlet69}). We find that the ME predictions for
thermodynamic variables and for the radial distribution function are
considerable improvements over the standard Bogoliubov variational result,
and are comparable to the perturbative results \cite%
{BarkerHenderson76} \cite{WCA}. (For a recent discussion of some of the
strengths and limitations of the perturbative approach see \cite{Germain02}.).
Despite the shortcomings of perturbation theory it remains
very popular because it provides quantitative insights at a much lower
computational cost than dynamical simulations. For recent applications to
the glass transition and other more complex systems see \cite{Lopez06} - \cite{Betancout08}.  
Although in this work the ME is not applied to such complex problems one may fully
expect that the information theory based ME method will yield insights not
only the thermodynamic behavior of complex systems but also about the
approximation methods needed to analyze them. Finally, our conclusions and
some remarks on further improvements are presented in section ~\ref{sec7}.

\section{General formalism of ME optimal approximations}

\label{sec2} 
Consider a system with microstates
labeled by $q$ (for example, the location in phase space or perhaps the
values of spin variables). Let the probability that the microstate lies
within a particular range $dq$ be given by the intractable canonical distribution 
\begin{equation}
P(q)dq=\frac{e^{-\beta H(q)}}{Z}dq,  \label{P(q)}
\end{equation}%
where $
Z=\int dq\,e^{-\beta H(q)}\stackrel{\textnormal{def}}{=}e^{-\beta F}
$
is partition function, in which $H(q)$ is the Hamiltonian and free energy $F$ is defined 
through this partition function. The goal is to generate an approximation $p(q)$
that is optimal in the sense that is ``closest'' to the ``exact'' 
distribution $P(q)$. It includes two steps. The first is to 
select a preferred trial distribution from a family of trials $p'(q|\theta)$,
each member of the trial family being
labeled by one or more parameters $\theta $. More generally, one could
define the trial family in a non-parametric way by specifying various
constraints. The second is to marginalize over parameters $\theta $ to 
obtain an optimal distribution.

\subsection{Entropic criterion for selecting preferred tractable PDFs}

\label{sec2-1} \noindent \textbf{Relative entropy as the selection criterion.} The selection of the preferred approximation is achieved by
ranking the distributions $p(q)$ according to increasing \emph{preference},
a real number $	\mathcal{D}[p]$ which we call the ``entropy'' of $p$. 
The numbers $\mathcal{D}[p]$ are such that if $p_{1}
$ is preferred over $p_{2}$, then $\mathcal{D}[p_{1}]>\mathcal{D}[p_{2}]$. Thus, by design, the
``preferred'' approximation $p$ is that which
maximizes the entropy $\mathcal{D}[p]$.

Next we determine the functional form of $\mathcal{D}[p]$. This is the general rule
that provides the criterion for preference; in our case it defines what we
mean by the ``closest'' or
``preferred'' approximation. The basic strategy 
\cite{Skilling88} is one of induction: (1) if a general rule exists, then it
must apply to special cases; (2) if in a certain special case we know which
is the best approximation, then this knowledge can be used to constrain the
form of $\mathcal{D}[p]$; and finally, (3) if enough special cases are known, then $%
\mathcal{D}[p]$ will be completely determined.

The known special cases are called the \textquotedblleft
axioms\textquotedblright\ of ME and they reflect the conviction that
whatever information was originally codified into the exact $P(q)$ is
important and should be preserved. The selected trial distribution should
coincide with the exact one as closely as possible and one should only
tolerate those minimal changes that are demanded by the information that
defines the family of trials. Three axioms and their consequences are listed
below. Detailed proofs and more extensive comments are given in \cite%
{Caticha03} and \cite{Caticha03b}.

\textbf{Axiom 1: Locality}. \emph{Local information has local effects.} If
the constraints that define the trial family do not refer to a certain
domain $D$ of the variable $q$, then the conditional probabilities $p(q|D)$
need not be revised, $p(q|D)=P(q|D)$. The consequence of the axiom is that
non-overlapping domains of $q$ contribute additively to the entropy: $%
\mathcal{D}[p]=\int dq\,F(p(q))$ where $F$ is some unknown function.

\textbf{Axiom 2: Coordinate invariance.} \emph{The ranking should not depend
on the system of coordinates. }The coordinates that label the points $q$ are
arbitrary; they carry no information. The consequence of this axiom is that $%
\mathcal{D}[p]=\int dq\,p(q)f(p(q)/m(q))$ involves coordinate invariants such as $%
dq\,p(q)$ and $p(q)/m(q)$, where the function $m(q)$ is a density, and both
functions $m$ and $f$ are, at this point, unknown.

Next we make a second use of Axiom 1 (locality). When there are no
constraints at all and the family of trials includes the exact $P(q)$ the
selected trial should coincide with $P(q)$; that is, the best approximation
to $P(q)$ is $P(q)$ itself. The consequence is that up to normalization the
previously unknown density $m(q)$ is the exact distribution $P(q)$.

\textbf{Axiom 3:\ Consistency for independent subsystems}. \emph{When a
system is composed of subsystems that are independent it should not matter
whether the approximation procedure treats them separately or jointly.}
Specifically, if $q=(q_{1},q_{2})$, and the exact distributions for the
subsystems, $P_{1}(q_{1})$ and $P_{2}(q_{2})$, are respectively approximated
by $p_{1}(q_{1})$ and $p_{2}(q_{2})$, then the exact distribution for the
whole system $P_{1}(q_{1})P_{2}(q_{2})$ should be approximated by $%
p_{1}(q_{1})p_{2}(q_{2})$. This axiom restricts the function $f$ to be a
logarithm.

The overall consequence of these axioms is that the trial approximations $%
p(q)$ should be ranked relative to the exact $P(q)$ according to their
(relative) entropy, 
\begin{equation}
\mathcal{D}[p|P]=-\int dq\,p(q)\log \frac{p(q)}{P(q)}.  \label{S[p|P]}
\end{equation}%
The derivation has singled out the relative entropy $\mathcal{D}[p|P]$ as \emph{the
unique functional to be used for the purpose of selecting a preferred
approximation}. Other functionals, may be useful for other purposes, but
they are not a generalization from the simple cases described in the axioms
above.

\noindent \textbf{Remark.} Suppose a member of a family of trial canonical
distributions $p'(q|\theta)$ with Hamiltonian $H(q|\theta)$ that are conditional 
probability distributions and depend on parameters 
$\theta =\{\theta ^{1},\ldots ,\theta ^{n}\}$, are given by
\begin{equation}
p'(q|\theta)dq=\frac{e^{-\beta H(q|\theta)}}{Z_{\theta }}dq,
\label{ptheta(q)}
\end{equation}%
where $
Z_{\theta }=\int dq\,e^{-\beta H(q|\theta)}\stackrel{\textnormal{def}}{=}e^{-\beta F_{\theta }}$, 
in which free energy $F_{\theta }$ is also defined.
The preferred trial is then selected by maximizing $\mathcal{D}[p'|P]$.
Substituting Eq. (\ref{P(q)}) and Eq. (\ref{ptheta(q)}) into Eq. (\ref{S[p|P]}) gives, 
\begin{equation}
\mathcal{D}[p'|P]=\beta \left( \langle H_{\theta }-H\rangle _{\theta
}-F_{\theta }+F\right) \,,  \label{S[ptheta|P]}
\end{equation}%
where $\langle \ldots \rangle _{\theta }$ refers to averages computed with
the distribution $p(q|\theta)$. The inequality $\mathcal{D}[p'|P]\leq 0$,
can then be written as 
\begin{equation}
F\leq F_{\theta }+\langle H-H_{\theta }\rangle _{\theta }\,.
\label{F_U-general}
\end{equation}%
Thus, maximizing $\mathcal{D}[p'|P]$ is equivalent to minimizing the quantity 
$F_{\theta }+\langle H-H_{\theta }\rangle _{\theta }$. This form of the
variational principle and its use to generate approximations is well known.
It is usually associated with the name of Bogoliubov \cite{Callen85} and it
is the main technique to generate mean field approximations for discrete
systems of spins on a lattice. 

\subsection{Marginalization for optimal PDF}

\label{sec2-3} The extent to which the preferred $\theta $ is preferred over
other values (\cite{Caticha03}, \cite{Caticha00}) is expressed
by the probability of $\theta $, $p(\theta )$. The original ME problem of
assigning a probability to $q$ is now broadened into assigning probabilities
to both $q$ and $\theta $.\ In this section we use ME again to find the preferred
joint distribution $p_{J}(q,\theta )=p(\theta)p(q|\theta)$. 
Note that this is the kind of problem where the Bayesian interpretation of
probabilities is essential. Within a frequentist interpretation it makes no
sense to talk about $p(\theta )$ or about $p(q|\theta )$ because $\theta $
is not a random variable; the value of $\theta $ is unknown but it is not
random.

To proceed we must ask a question. What is the prior
distribution, that is, what do we know about $q$ and $\theta $ before the
trial family is specified? 

The joint prior $m(q,\theta )$ can be expressed according to the 
product rule as, $m(q,\theta )=m(q|\theta)m(\theta )$, where $m(q|\theta)$ 
is conditional probability of observing system in state $q$ given parameter 
$\theta$. Our goal is to determine the preferred $p_{J}(q,\theta )$ that 
is closest to the prior $m(q,\theta )$  that reflects our initial knowledge about $q$ and 
ignorance about the $\theta $s. Initially we know 
nothing about $\theta $, not even how it is related to $q$. 
The prior that represents this state of knowledge is a product%
\begin{equation}
m(q,\theta )=P(q)\mu (\theta )~.
\end{equation}%
Indeed, when $m(q,\theta )$ is a product no correlations between $\theta $
and $q$ are introduced which means that information about $q$ tells us
nothing about $\theta $ and vice versa. The first factor in $m(q,\theta )$
reflects our prior knowledge about $q$: the distribution for $q$ is known to
be the exact $P(q)$. The second factor reflects our complete ignorance about 
$\theta $: we choose $\mu (\theta )$ to be as uniform as possible. Our
method applies whether $\theta $ is a discrete or a continuous variable.
When $\theta $ is a continuous variable. Then the uniform distribution $\mu
(\theta )$ is such that makes equal volumes in $\theta $ space equally
likely. To define these volumes we apply method of information 
geometry \cite{Cencov81} and note that distances in $%
\theta $-space are uniquely defined because the $\theta $s are labels on
probability distributions. Cases where $\theta $ is a discrete variable are simpler.
The relevant entropies involve sums over $\theta _{i}$ rather than integrals
and the natural uniform distribution is $\mu (\theta _{i})=\rm{constant}$.
In what follows we concentrate on the more challenging continuous
case. The unique distance between $\theta $ and $\theta
+d\theta $ is given by the Fisher-Rao metric \cite{Cencov81}, $d\ell
^{2}=\gamma _{ij}d\theta ^{i}d\theta ^{j}$, where 
\begin{equation}
\gamma _{ij}=\int dq\,p(q|\theta)\frac{\partial \log ~p(q|\theta)}{%
\partial \theta ^{i}}\frac{\partial \log ~p(q|\theta)}{\partial \theta
^{j}}.  \label{Fisher metric}
\end{equation}%
Accordingly, the volume of a small region $d\theta $ is $\gamma ^{1/2}(\theta )d\theta $,
where $\gamma (\theta )$ is the determinant of $\gamma _{ij}$. Up to an
irrelevant normalization, the distribution $\mu (\theta )$ that is uniform
in $\theta $ is given by $\mu (\theta )=\gamma ^{1/2}(\theta )$.

The preferred approximation $p_{J}(q,\theta )$ to the joint distribution $%
P(q)\gamma ^{1/2}(\theta )$ is then obtained maximizing the entropy%
\begin{equation}
\mathcal{D}[p_{J}|\gamma ^{1/2}P]=-\int dq\,d\theta \,p(\theta )p_{\theta }(q)\,\log 
\frac{p(\theta )p_{\theta }(q)}{\gamma ^{1/2}(\theta )P(q)},
\label{sigma[p]}
\end{equation}%
by varying $p(\theta )$ subject to $\int d\theta \,p(\theta )=1$. The final
result for the probability that $\theta $ lies within the small volume $%
\gamma ^{1/2}(\theta )d\theta $ is 
\begin{equation}
p(\theta )d\theta =\frac{1}{\zeta }\,\,e^{\mathcal{D}[p_{\theta }|P]}\gamma
^{1/2}(\theta )d\theta ,  \label{p(theta)}
\end{equation}%
where $\mathcal{D}[p_{\theta }|P]$ is given in Eq. (\ref{S[ptheta|P]}) and $\zeta $ is a
normalization constant. Note also that the density $\exp \mathcal{D}[p_{\theta }|P]$
is a scalar function and the presence of the Jacobian factor $\gamma
^{1/2}(\theta )$ makes Eq. (\ref{p(theta)}) manifestly invariant under changes of
the coordinates $\theta $. Eq. (\ref{p(theta)}) expresses the degree to which
values of $\theta $ away from the preferred value are ruled out; it tells us
that the preferred value of $\theta $ is that which maximizes the
probability density $\exp \mathcal{D}[p_{\theta }|P]$.

Finally, now that we have determined the preferred joint distribution $%
p_{J}(q,\theta )=p(\theta )p(q|\theta)$ we can marginalize $\theta $ and
use the average 
\begin{equation}
\bar{p}(q)=\int d\theta \,p(\theta )p(q|\theta)  \label{pbar}
\end{equation}%
as the best approximation we can construct out of the given trial family.
This approximation is expected to be better than any individual $p(q|\theta)$ 
for the same reason that the mean is expected to be a better estimator
than the mode -- it minimizes the variance.

This concludes the first part of our paper. To summarize: our main results
consist in the justification of the relative entropy Eq. (\ref{S[p|P]}) as the
uniquely natural functional to select the preferred approximations and the
derivation of a quantitative measure of the degree to which the various
trials are preferred, Eq. (\ref{p(theta)}). The final result for the best
approximation is Eq. (\ref{pbar}).

Next we illustrate how this ME formalism is used in a specific
example: simple fluids.

\section{ME optimal hard-sphere approximation for simple fluids}

\label{sec3}

\subsection{Basic features of simple fluids} 
A simple fluid composed of $N$ single atom molecules is described by the
Hamiltonian 
\begin{equation}
H(q_{N})=\sum\limits_{i=1}^{N}\,\frac{p_{i}^{2}}{2m}+U \quad \mathrm{ with } \quad
U=\sum\limits_{i>j}^{N}u(r_{ij})~,  \label{Actual-H}
\end{equation}%
where $q_{N}=\{p_{i},r_{i};\;i=1,...,N\}$ and the many-body interactions are
approximated by a pair interaction, $u(r_{ij})$ where $r_{ij}=\left\vert
r_{i}-r_{j}\right\vert $. The probability that the positions and momenta of
the molecules lie within the phase space volume 
\begin{equation}
dq_{N}=\frac{1}{N!h^{3N}}\prod\limits_{i=1}^{N}d^{3}p_{i}d^{3}r_{i}
\end{equation}
is given by canonical distribution 
\begin{equation}
P_{f}(q_{N})\,dq_{N}=\frac{1}{Z_{f}}e^{-\beta H(q_{N})\,}\,dq_{N},
\label{exact dist}
\end{equation}%
where $Z_{f}=\int dq_{N}\,e^{-\beta H(q_{N})}$.
For fluids dominated by pair interactions most thermodynamic quantities of
interest can be written in terms of the one- and two-particle density
distributions 
\begin{equation}
n(r)=\langle \hat{n}(r)\rangle \quad \mathrm{and} \quad n^{\left(%
2\right) }(r_{1},r_{2})=\langle \hat{n}^{\left( 2\right)
}(r_{1},r_{2})\rangle  \label{numberdensity}
\end{equation}%
where 
\begin{equation}
\hat{n}(r)=\sum\limits_{i}\,\delta (r-r_{i})
\end{equation}%
and 
\begin{equation}
\hat{n}^{\left( 2\right) }(r_{1},r_{2})=\sum\limits_{i,j(i\neq j)}\,\delta
(r_{1}-r_{i})\,\delta (r_{2}-r_{j})\;.
\end{equation}%
The two-particle correlation function, 
\begin{equation}
g(r_{1},r_{2})=\frac{n^{\left( 2\right) }(r_{1},r_{2})}{n(r_{1})n(r_{2})}~,
\label{rdf-general}
\end{equation}%
measures the extent to which the structure of liquids deviates from complete
randomness. If the fluid is homogeneous and isotropic $n(r)=\rho =N/V$ and $%
g(r_{1},r_{2})=g(|r_{1}-r_{2}|)=g(r)$ where $\rho $ is the bulk density and $%
g(r)$ is the radial distribution function (RDF). Then, the pressure is given
by 
\begin{equation}
\frac{PV}{Nk_{B}T}=1-\frac{\beta \rho }{6}\int d^{3}r\,r\frac{du\left(
r\right) }{dr}g\left( r\right) \,,  \label{eq of state}
\end{equation}%
where $\beta \overset{\mathrm{def}}{=}1/k_{B}T$ \cite{BarkerHenderson76}-%
\cite{Kalikmanov02}.

\subsection{Hard-sphere approximation} To account for the short-distance repulsion we consider a family of trials
composed by distributions that describe a gas of hard spheres of diameter $%
r_{d}$. For each $r_{d}$ the Hamiltonian is 
\begin{equation}
H_{hs}(q_{N}\left\vert r_{d}\right. )=\sum\limits_{i=1}^{N}\,\frac{p_{i}^{2}%
}{2m}+U_{hs}  \label{Hhs}
\end{equation}%
with 
\begin{equation}
U_{hs}=\sum\limits_{i>j}^{N}u_{hs}(r_{ij}|r_{d})~,
\end{equation}%
where 
\begin{equation}
u_{hs}(r\left\vert r_{d}\right. )=\left\{ 
\begin{array}{ccc}
0 & \mathrm{for} & r\geq r_{d} \\ 
\infty & \mathrm{for} & r<r_{d}%
\end{array}%
\right.
\end{equation}%
and the corresponding probability distribution is 
\begin{equation}
P_{hs}(q_{N}\left\vert r_{d}\right. )=\frac{1}{Z_{hs}}e^{-\beta
H_{hs}(q_{N}\left\vert r_{d}\right. )}\,.  \label{Phs}
\end{equation}%
The partition function and the free energy $F_{hs}(T,V,N\left\vert
r_{d}\right) $ are $
Z_{hs}$ $=\int dq_{N}\,\ e^{-\beta H_{hs}(q_{N}\left\vert r_{d}\right. )}\,%
$ $\overset{\mathrm{def}}{=}e^{-\beta F_{hs}(T,V,N\left\vert r_{d}\right. )}\,.$
Two objections that can be raised for choosing $P_{hs}(q_{N}|r_{d})$ as
trials are, first, that they do not take the long-range interactions into
account; and second, that the actual short range potential is not that of
hard spheres. These are points to which we will return later. A third
objection, and this is considerably more serious, is that the exact
hard-sphere RDF is not known. However, it can be calculated within the
approximation of Percus and Yevick (PY) for which there exists an exact
analytical solution (\cite{PercusYevick58}, \cite{Percus62}, \cite{Wertheim63}) which is
reasonably simple and in good agreement with numerical simulations over an
extended range of temperatures and densities, except perhaps at high
densities. There are several successful proposals \cite{Bravo91} to improve
upon the PY RDF but they also represent an additional level of complication.
The simpler PY RDF is sufficiently accurate for our current objective -- to
illustrate the application and study the broad features of the ME approach.

The PY RDF can be written in terms of the Laplace transform of $%
rg_{hs}(r\left\vert r_{d}\right. )$ \cite{Wertheim63}, 
\begin{equation}
G(s) =\int\limits_{0}^{\infty }dy~yg_{hs}(yr_{d}|r_{d})e^{-sy}
=\frac{sL(s)}{12\eta \left[ L(s)+M(s)e^{s}\right] },  \label{G(s)}
\end{equation}%
where $y$ is a dimensionless variable $y=r/r_{d}$, 
\begin{equation}
L(s)=12\eta \left[ \left( 1+\frac{1}{2}\eta \right) s+\left( 1+2\eta \right) %
\right] ,
\end{equation}%
\begin{equation}
M(s) =\left( 1-\eta \right) ^{2}s^{3}+6\eta \left( 1-\eta \right) s^{2} 
+18\eta ^{2}s-12\eta \left( 1+2\eta \right) ,
\end{equation}%
and $\eta $ is the packing fraction, 
\begin{equation}
\eta \overset{\mathrm{def}}{=}\frac{1}{6}\pi \rho r_{d}^{3}\quad \rm{with}%
\quad \rho =\frac{N}{V}~.  \label{eta}
\end{equation}%
The RDF $g_{hs}(r\left\vert r_{d}\right. )$ is obtained from the inverse
transform using residues \cite{Throop65}.

The equation of state can then be computed in two alternative ways, either
from the ``pressure'' equation or from the ``compressibility'' equation but,
since the result above for $g_{hs}(r\left\vert r_{d}\right. )$ is not exact,
the two results do not agree. It has been found that better agreement with
simulations and with virial coefficients is obtained taking an average of
the two results with weights 1/3 and 2/3 respectively. The result is the
Carnahan-Starling equation of state, \cite{BarkerHenderson76}-\cite%
{Kalikmanov02} 
\begin{equation}
\left( \frac{PV}{Nk_{B}T}\right) _{hs}=\frac{1+\eta +\eta ^{2}-\eta ^{3}}{%
\left( 1-\eta \right) ^{3}}.  \label{HS-P}
\end{equation}%
The free energy, derived by integrating the equation of state, is 
\begin{equation}
F_{hs}(T,V,N\left\vert r_{d}\right. )=Nk_{B}T\left[ -1+\ln \rho \Lambda ^{3}+%
\frac{4\eta -3\eta ^{2}}{\left( 1-\eta \right) ^{2}}\right] ,\mathrm{\ }
\label{HS-Free energy}
\end{equation}%
where $\Lambda =(2\pi \hbar ^{2}/mk_{B}T)^{1/2}$, and the entropy is 
\begin{equation}
\mathcal{D}_{hs}=-\left( \frac{\partial F_{hs}}{\partial T}\right) _{N,V}=\frac{F_{hs}%
}{T}+\frac{3}{2}Nk_{B}.  \label{HS-S}
\end{equation}%
It must be remembered that these expressions are not exact. They are
reasonable approximations for all densities up to almost crystalline
densities (about $\eta \approx 0.5$). However, they fail to predict the
face-centered-cubic phase when $\eta $ is in the range from $0.5$ up the
close-packing value of $0.74$.

\section{The ME formalism}
\subsection{Preferred hard-sphere PDF}
\label{sec4} As discussed in section ~\ref{sec2}, the trial $%
P_{hs}(q_{N}|r_{d})$ that is \textquotedblleft closest\textquotedblright\ to
the \textquotedblleft exact\textquotedblright\ $P_{f}(q_{N})$ is found by
maximizing the relative entropy $\mathcal{D}\left[ p|P\right] $, Eq. (\ref{S[p|P]}), with $%
p=P_{hs}(q_{N}|r_{d})$ given by Eq. (\ref{Phs}) and $P=P_{f}(q_{N})$ given by %
Eq. (\ref{exact dist}). According to Eq. (\ref{F_U-general}), it is equivalent to
minimize 
\begin{equation}
F_{U}\overset{\mathrm{def}}{=}F_{hs}+\langle U-U_{hs}\rangle _{hs}
\end{equation}%
over all diameters $r_{d}$, where $\langle \cdots \rangle _{hs}$ is computed
with $P_{hs}(q_{N}|r_{d})$. Thus, the variational approximation to the free
energy is 
\begin{equation}
F\left( T,V,N\right) \approx F_{U}(T,V,N\left\vert r_{m}\right. )\overset{%
\mathrm{def}}{=}\min_{r_{d}}~F_{U}(T,V,N\left\vert r_{d}\right. )\,,
\label{mini-Fu}
\end{equation}%
where $r_{m}$ is the preferred diameter.

To calculate $F_{U}$ use 
\begin{equation*}
\langle U-U_{hs}\rangle _{hs}=\frac{1}{2}\int d^{3}rd^{3}r^{\prime }\mathrm{%
\ }n_{hs}^{\left( 2\right) }(r,r^{\prime })\left[ u(r-r^{\prime
})-u_{hs}(r-r^{\prime }|r_{d})\right] \mathrm{~,}
\end{equation*}%
where $n_{hs}^{\left( 2\right) }(r,r^{\prime })=\langle \hat{n}^{\left(
2\right) }(r,r^{\prime })\rangle _{hs}$. But $u_{hs}(r-r^{\prime }|r_{d})=0$
for $\left\vert r-r^{\prime }\right\vert \geq r_{d}$ while $n_{hs}^{\left(
2\right) }(r,r^{\prime })=0$ for $\left\vert r-r^{\prime }\right\vert \leq
r_{d}$, therefore 
\begin{equation}
F_{U}=F_{hs}+\langle U\rangle _{hs}  \label{F_U}
\end{equation}%
with 
\begin{equation}
\langle U\rangle _{hs}=\frac{1}{2}N\rho \int d^{3}r\,u(r)g_{hs}(r\left\vert
r_{d}\right. ),
\end{equation}%
where we have assumed that the fluid is isotropic and homogeneous, $%
n_{hs}^{\left( 2\right) }(r,r^{\prime })=n_{hs}^{\left( 2\right)
}(\left\vert r-r^{\prime }\right\vert )$, and introduced the hard-sphere RDF
\begin{equation}
g_{hs}(r\left\vert r_{d}\right. )\overset{\mathrm{def}}{=}\frac{%
n_{hs}^{\left( 2\right) }(r)}{\rho ^{2}}.  \label{rdf-hs}
\end{equation}%
Note that the approximation does not consist of merely replacing the exact
free energy $F$ by a hard-sphere free energy $F_{hs}$ which neglects the
effects of long range attraction; $F$ is approximated by $F_{U}(r_{m})$
which includes attraction effects through the $\langle U\rangle _{hs}$ term
in Eq. (\ref{F_U}). This addresses the first of the two objections mentioned
earlier: the real fluid with interactions given by $u$ is not being replaced
by a hard-sphere fluid. The internal energy is approximated by $\langle
H\rangle _{hs}=\frac{3}{2}Nk_{B}T+\langle U\rangle _{hs}$ and not by $%
\langle H_{hs}\rangle _{hs}=\frac{3}{2}Nk_{B}T$.

To calculate $\langle U\rangle _{hs}$ it is convenient to write it in terms
of $V(s)$, the inverse Laplace transform of $ru(r)$, 
\begin{equation}
yu(yr_{d})=\int\limits_{0}^{\infty }ds~V(s)e^{-sy}.
\end{equation}%
For example, for a Lennard-Jones potential, 
\begin{equation}
u(r)=4\varepsilon \left[ \left( \frac{\sigma }{r}\right) ^{12}-\left( \frac{%
\sigma }{r}\right) ^{6}\right] ,  \label{LJ Pot}
\end{equation}%
we have 
\begin{equation}
V(s)=4\varepsilon \left[ \left( \frac{\sigma }{r_{d}}\right) ^{12}\frac{%
s^{10}}{10!}-\left( \frac{\sigma }{r_{d}}\right) ^{6}\frac{s^{4}}{4!}\right]
.
\end{equation}%
Then, using equations Eq. (\ref{F_U}) and (\ref{G(s)}) gives 
\begin{equation}
\langle U\rangle _{hs}=12N\eta \int\limits_{0}^{\infty }ds\mathrm{~}V(s)G(s).
\label{F_U-LT}
\end{equation}%
Finally, it remains to minimize $F_{U}$ in Eq. (\ref{F_U}) to determine the
preferred diameter $r_{m}$. This is done numerically in an explicit example for argon
in section ~\ref{sec6}.\ 

\subsection{Marginalization for optimal hard-sphere PDF}

\label{sec5} The ME method as pursued in the last section has led us to
determine a preferred hard-sphere diameter. It fails to take the
softness of the repulsive core into account. As discussed in section ~\ref%
{sec2-3}, our best assessment of
the distribution of $q_{N}$ is given by the marginal over $r_{d}$, 
\begin{equation}
\bar{P}_{hs}(q_{N}) \overset{\mathrm{def}}=\int dr_{d}\mathrm{\ }%
P_{J}(q_{N},r_{d})  
=\int dr_{d}\mathrm{~}P_{d}(r_{d})P_{hs}(q_{N}|r_{d})\mathrm{.{\ }}
\label{Pbar(q)}
\end{equation}
The corresponding best
approximation to the RDF is obtained using %
Eq. (\ref{numberdensity}), (\ref{rdf-general}), and (\ref{rdf-hs})
\begin{equation}
\bar{g}_{hs}(r)=\int dq_{N} \bar{P}_{hs}(q_{N}) \hat{n}^{2}(r)/\rho^{2}
=\int dr_{d}\mathrm{~}P_{d}(r_{d})g_{hs}(r\left\vert
r_{d}\right. )\ .  \label{ghs bar}
\end{equation}
By averaging over all hard-sphere diameters we are effectively describing a
soft-core potential. Since $\bar{g}_{hs}(r)$ takes into account soft-core effects while $%
g_{hs}(r\left\vert r_{m}\right. )$ does not, we expect that it will lead to
improved estimates for all other thermodynamic quantities. 

However, we should emphasize that the distribution over
the hard-sphere diameters $P_{d}(r_{d})$ is not being introduced in an ad
hoc way in order to \textquotedblleft fix\textquotedblright\ the variational
method. The introduction of $P_{d}(r_{d})$ is mandated by the ME method
(section ~\ref{sec2-3}). The distribution of diameters is given by %
Eq. (\ref{p(theta)})
\begin{equation}
P_{d}(r_{d})dr_{d}=\frac{e^{\mathcal{D}\left[ P_{hs}|P\right] }}{\zeta }\gamma
^{1/2}\left( r_{d}\right) dr_{d}=\frac{e^{-\beta F_{U}}}{\zeta _{U}}\gamma
^{1/2}\left( r_{d}\right) dr_{d},  \label{Pd(rd)}
\end{equation}%
where $\mathcal{D}\left[ P_{hs}|P\right] =\beta \left( F-F_{U}\right) $, the partition
functions $\zeta $ and $\zeta _{U}$ are given by 
\begin{equation}
\zeta =e^{\beta F}\zeta _{U}\quad \rm{with}\mathrm{\quad }\zeta _{U}=\int
dr_{d}\ \gamma ^{1/2}\left( r_{d}\right) e^{-\beta F_{U}},
\end{equation}%
and the natural distance $d\ell ^{2}=\gamma (r_{d})dr_{d}^{2}$ in the space
of $r_{d}$s is given by the Fisher-Rao metric, 
\begin{equation}
\gamma (r_{d})=\int dq_{N}\,P_{hs}(q_{N}\left\vert r_{d}\right. )\left( 
\frac{\partial \log P_{hs}(q_{N}\left\vert r_{d}\right. )}{\partial r_{d}}%
\right) ^{2}.  \label{FR-metric}
\end{equation}

A convenient way to calculate the Fisher-Rao metric is to express it as a
second derivative of the entropy Eq. (\ref{S[p|P]}) of $p=P_{hs}(q_{N}\left\vert
r_{d}^{\prime }\right. )$ relative to $P=P_{hs}(q_{N}\left\vert r_{d}\right.
)$,
\begin{equation}
\gamma (r_{d})=-\left. \frac{\partial ^{2}}{\partial r_{d}^{\prime 2}}\mathcal{D}\left[
P_{hs}(\cdot \left\vert r_{d}^{\prime }\right. )\left\vert P_{hs}(\cdot
\left\vert r_{d}\right. )\right. \right] \right\vert _{r_{d}^{\prime }=r_{d}} ,
\label{det g}
\end{equation}%
where
\begin{equation}
\mathcal{D}\left[ P_{hs}(\cdot |r_{d}^{\prime })|P_{hs}(\cdot \left\vert r_{d}\right. )%
\right] =\beta \left[ \left. F_{hs}\right\vert _{r_{d}^{\prime
}}^{r_{d}}-\langle \left. U_{hs}\right\vert _{r_{d}^{\prime
}}^{r_{d}}\rangle _{r_{d}^{\prime }}\right] ,
\end{equation}%
and $\langle \cdots \rangle _{r_{d}^{\prime }}$ is the average over $%
P_{hs}(q_{N}|r_{d}^{\prime })$. As we argued above Eq. (\ref{F_U}) the
expectation of the potential energy $\langle U_{hs}\left( r_{d}^{\prime
}\right) \rangle _{r_{d}^{\prime }}$ vanishes because the product $%
u_{hs}(r\left\vert r_{d}^{\prime }\right. )g_{hs}(r\left\vert r_{d}^{\prime
}\right. )$ vanishes for both $r<r_{d}^{\prime }$ and $r>r_{d}^{\prime }$.
Similarly, $\langle U_{hs}\left( r_{d}\right) \rangle _{r_{d}^{\prime }}=0$
when $r_{d}^{\prime }>r_{d}$. However, when $r_{d}^{\prime }<r_{d}$ the
expectation $\langle U_{hs}\left( r_{d}\right) \rangle _{r_{d}^{\prime }}$
diverges. The divergence is a consequence of the unphysical nature of the
hard-sphere model. For more realistic continuous potentials the distance
between $r_{d}^{\prime }=r_{d}+dr_{d}$ and $r_{d}$ is the same as the
distance between $r_{d}^{\prime }=r_{d}-dr_{d}$ and $r_{d}$. We can then
always choose $r_{d}^{\prime }\geq r_{d}$ and define $\gamma (r_{d})$ in %
Eq. (\ref{det g}) as the limit $r_{d}^{\prime }=r_{d}+0^{+}$. Then, using %
Eq. (\ref{HS-Free energy}) for $F_{hs}$, we have 
\begin{equation}
\gamma (r_{d})=\beta \left. \frac{\partial ^{2}F_{hs}\left( r_{d}^{\prime
}\right) }{\partial r_{d}^{\prime 2}}\right\vert _{r_{d}^{\prime
}=r_{d}+0^{+}}\mathrm{\ } 
=N\pi \rho r_{d}\frac{4+9\eta -4\eta ^{2}}{\left( 1-\eta \right) ^{4}}~.
\label{det g-exact}
\end{equation}

To summarize, the distribution of diameters $P_{d}(r_{d})$ is given by %
Eq. (\ref{Pd(rd)}) with $F_{U}$ given by Equations (\ref{F_U}), (\ref{HS-Free
energy}), (\ref{F_U-LT}) and $\gamma $ given by Eq. (\ref{det g-exact}). Our best
approximation to the ``exact'' $P(q_{N})$
is the $\bar{P}_{hs}(q_{N})$ given in Eq. (\ref{Pbar(q)}). However, there is
a problem. Since the free energy $F_{U}$ is an extensive quantity, $%
F_{U}\propto N$, for large $N$ the distribution $P_{d}(r_{d})\sim \exp
-\beta F_{U}$ is very sharply peaked at the preferred diameter $r_{m}$. When
choosing a \emph{single} preferred diameter for a macroscopic fluid sample we
find that ME confers overwhelming probability to the preferred value. This is
not surprising. The same thing happens when we calculate the global
temperature or density of a macroscopic sample and yet local fluctuations
can be important. The question then, is whether local fluctuations are
relevant to the particular quantities we want to calculate. We argue that
they are.

From the very definition of $g(r)$ as the probability that given an atom at
a certain place another atom will be found at a distance $r$, it is clear
that $g(r)$ refers to purely local behavior and should be influenced by
local fluctuations. To the extent that the preferred diameter $r_{m}$ depends
on temperature and density we expect that local temperature and/or density
fluctuations would also induce local diameter fluctuations.

For the purpose of calculating $g(r)$ the system is effectively reduced to
the small number of atoms $N_{\mathrm{eff}}$ in the local vicinity of the
reference atom at the origin. In order to develop a systematic, fully ME method for the
determination of the effective number of particles $N_{\mathrm{eff}}$ that
are locally relevant, one needs to use trial probability distributions that 
allow inhomogeneities in the hard sphere diameters. Yet determination of such 
trial probability distributions requires further investigations.
Since we are interested in demonstrating ME formalism for optimizing approximations 
in this work only, we consider a rather simple approach to estimate $N_{\mathrm{eff}}$. 
Based on our ME approach to a mean field approximation for fluids 
\cite{Tseng02}, we have shown the RDF is given by 
\begin{equation}
g_{MF}\left( r\right)=e^{-\beta u\left( r\right)-%
\beta\rho\int d^{3}r'u(r')(g_{MF}(r-r')-1) } \mathrm{~.{\ }}  \label{estimaterule}
\end{equation}%
For a sufficiently dilute gas, the results of the mean field approximation  
are comparable with experimental results. The $g_{MF}\left( r\right)$ 
is approximately close to the true, exact RDF $g(R)$ for $r<\sigma$, the Lennard-Jones parameter. A fluid of hard spheres gives $g(r|r_{d})=0$ for $r<r_{d}$ and cannot
reproduce the behavior of Eq. (\ref{estimaterule}). However, once we recognize that
we can use a statistical mixture, Eq. (\ref{Pbar(q)}), we can tune the size $N_{%
\mathrm{eff}}$ of the cell and thereby change the width of $P_{d}\left(
r_{d}\right) $ so that the RDF $\bar{g}_{hs}\left( r\right) $ of Eq. (\ref{ghs bar}) 
reproduces the known short-distance behavior of Eq. (\ref{estimaterule}).

\section{Numerical demonstrations: Lennard-Jones ``argon''}

\label{sec6} One of the difficulties in testing theories about fluids
against experimental data is that it is not easy to see whether
discrepancies are to be blamed to a faulty approximation or to a wrong
intermolecular potential. This is why theories are normally tested against
molecular dynamics numerical simulations where there is control over the
intermolecular potential. In this section we compare ME results against
simulation results \cite{Verlet68} for a fluid of monoatomic molecules
interacting through a Lennard-Jones potential, Eq. (\ref{LJ Pot}). The parameters 
$\varepsilon $ and $\sigma $ (the depth of the well, $u|_{\min
}=-\varepsilon $, and the radius of the repulsive core, $u(\sigma )=0$,
respectively) are chosen to model argon: $\varepsilon =1.03\times 10^{-2}$ $%
\mathrm{eV}$ and $\sigma =3.405 ${\AA}. 

\subsection{Preliminary examinations}
\label{sec6-1} 
\begin{figure}[ht]
\centering
\includegraphics[width=5in,height=3.5in]{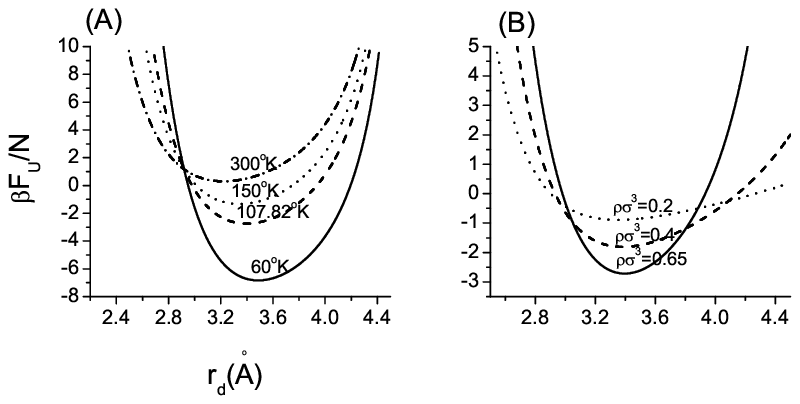} 
\caption{(A): The free energy $F_{U}$ as a function of hard-sphere diameter $%
r_{d}$ for argon at a density of $\protect\rho \protect\sigma ^{3}=0.65$ for
different temperatures. The best $r_{d}$ is that which minimizes $F_{U}$.
(B): $F_{U}$ as a function of $r_{d}$ for argon at $T=107.82$ $\mathrm{K}$
for different densities.}
\label{fig1}
\end{figure}

\noindent \textbf{The free energy $F_{U}$.}
Figure \ref{fig1}.(A) shows the free energy $F_{U}/Nk_{B}T$
as a function of hard-sphere diameter $r_{d}$ for argon at a fixed density
of $\rho \sigma ^{3}=0.65$ for different temperatures. Figure \ref{fig1}.(B)
shows $F_{U}/Nk_{B}T $ as a function of $r_{d}$ for several densities at
fixed $T=107.82$ $\mathrm{K}$. Since the critical point for argon is at $%
T_{c}=150.69$ $\mathrm{K}$ and $\rho _{c}\sigma ^{3}=0.33$ all these curves,
except that at $300$ $\mathrm{K}$, lie well within the liquid phase. The
increase of $F_{U}/Nk_{B}T$ for high values of $r_{d}$ is due to short range
repulsion between the hard spheres described by $F_{hs}/Nk_{B}T$. The
increase for low $r_{d}$ is due to the Lennard-Jones short-range repulsion
as described by $\langle U\rangle _{hs}/Nk_{B}T$.

The preferred $r_{d}$ is that which minimizes $F_{U}$ and depends both on
temperature and density. The preferred diameter decreases as the temperature
increases because atoms with higher energy can penetrate deeper into the
repulsive core. The dependence with density is less pronounced.

\begin{figure}[ht]
\centering
\includegraphics[width=5in,height=3.5in]{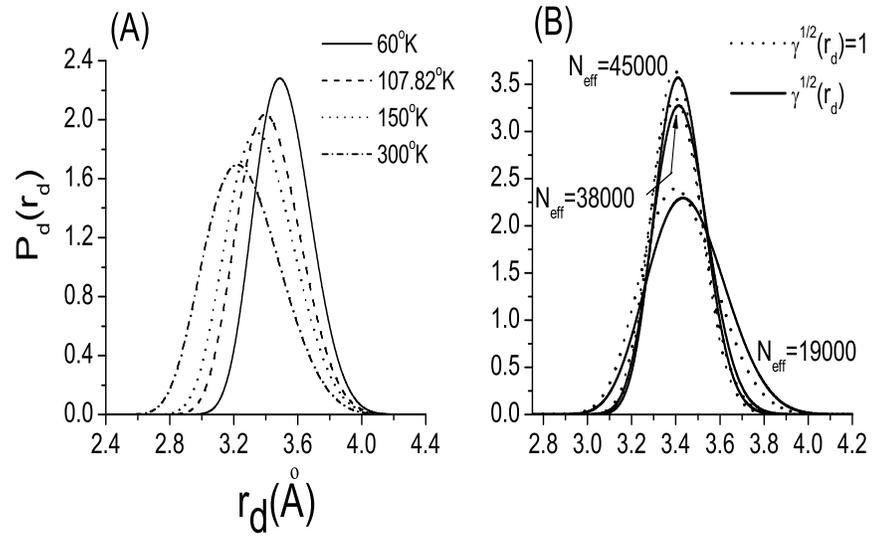} 
\caption{(A): The distribution of hard-sphere diameters $r_{d}$ for argon
for several temperatures at density $\protect\rho \protect\sigma ^{3}=0.65$
for $N_{\mathrm{eff}}=13500$. (B): $P_{d}\left( r_{d}\right) $ for various $%
N_{\mathrm{eff}}$ at $T=107.82$ $\mathrm{K}$ and $\protect\rho \protect%
\sigma ^{3}=0.65$. By setting $\protect\gamma ^{1/2}=1$ (dotted lines) we
see that the effect of the $\protect\gamma ^{1/2} $ factor is to cause a
slight shift of the distribution.}
\label{fig3}
\end{figure}

\noindent \textbf{The distribution of diameters $P_{d}(r_{d})$.}
In section ~\ref{sec5} we argued that the effective number of
molecules that is relevant to the local structure of the fluid is not the
total number of molecules in the system $N$, but a smaller number, $N_{%
\mathrm{eff}}$. In Fig \ref{fig3}.(A) we plot the distribution of diameters $%
P_{d}(r_{d})$ for different temperatures, for a fixed fluid density of $\rho
\sigma ^{3}=0.65$, and for an arbitrarily chosen $N_{\mathrm{eff}}=13500$.
As expected the distribution shifts to higher diameters as the temperature
decreases. Notice also that the distribution becomes narrower at lower
temperatures in agreement with the fact that a hard-sphere approximation is
better at low $T$ \cite{BarkerHenderson76}.

Figure \ref{fig3}.(B) shows that increasing $N_{\mathrm{eff}}$ (with fixed
density $\rho $) decreases the width of $P_{d}(r_{d})$ (solid lines) and
induces a slight shift of the whole distribution. This is due to the
dependence $\sim (N_{\mathrm{eff}}r_{d})^{1/2}$ of the Fisher-Rao measure $%
\gamma ^{1/2}\left( r_{d}\right) $ in \ref{det g-exact}. Figure \ref{fig3}%
.(B) also explores the influence of $\gamma ^{1/2}\left( r_{d}\right) $ by
comparing the actual distributions $P_{d}(r_{d})$ (solid lines) with the
distributions $e^{-\beta F_{U}\left( r_{d}\right) }$ (dotted lines) which
are obtained by setting $\gamma ^{1/2}=1$ in Eq. (\ref{Pd(rd)}). The effect of $%
\gamma ^{1/2}$ is to shift the distribution slightly to higher $r_{d}$.

\subsection{Two properties of argon}
\label{sec6-2} 
\begin{figure}[th]
\centering
\includegraphics[width=5.0in,height=4.0in]{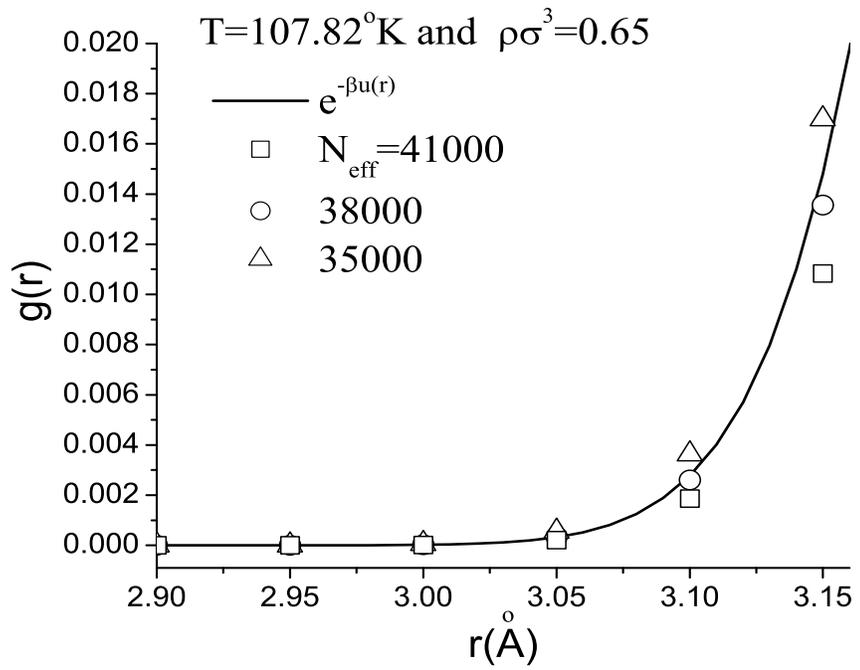}
\caption{ {Estimating $N_{\mathrm{eff}}$ by requiring that $\bar{g}_{hs}(r)$
have the correct short-distance behavior $e^{-\protect\beta u(r)}$.}}
\label{fig5}
\end{figure}

\noindent \textbf{The radial distribution function.}
We are finally ready to calculate the radial distribution $%
g(r)$ for argon. We start by estimating the number of molecules $N_{\mathrm{%
eff}}$ that are locally relevant; as explained earlier we choose $N_{\mathrm{%
eff}}$ so that our best approximation $\bar{g}_{hs}\left( r\right) $, %
Eq. (\ref{ghs bar}), reproduces the known short-distance behavior $g_{MF}\left( r\right)$, 
Eq. (\ref{estimaterule}), for $r\ll \sigma$. We have found that the estimates for $%
N_{\mathrm{eff}}$ need not be very accurate but that they must be obtained
for each value of the temperature and density. In Fig \ref{fig5} we show an
example of the short-distance behavior of $\bar{g}_{hs}$ for three values of 
$N_{\mathrm{eff}}$ at $T=107.82$ $\mathrm{K}$ and $\rho \sigma ^{3}=0.65$;
using a Chi-square fit in the range from $r=2.9$ to $3.1$ $\mathrm{{%
\mathring{A}}}$ the selected best value of $N_{\mathrm{eff}}$ is around
38000.

\begin{figure}[th]
\centering
\begin{picture}(0,150)(0,0)
\put(0,150){(A)}
\end{picture}
  \includegraphics[width=2.0in,height=2.0in]{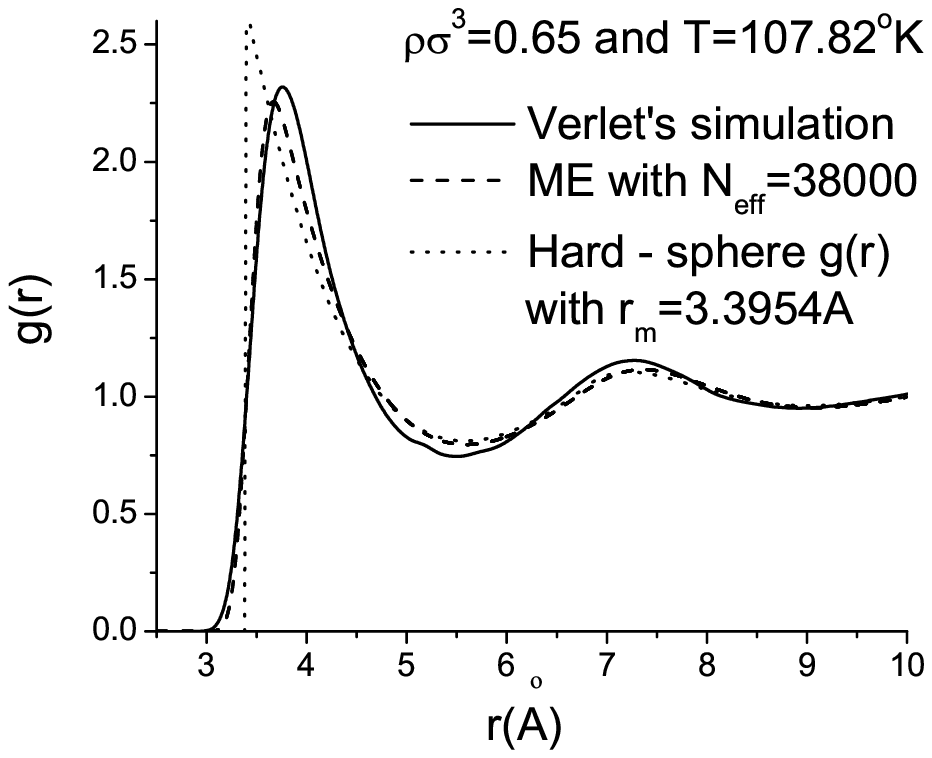}
\begin{picture}(4,150)(0,0)
\put(0,150){(B)}
\end{picture}
 \includegraphics[width=2.0in,height=2.0in]{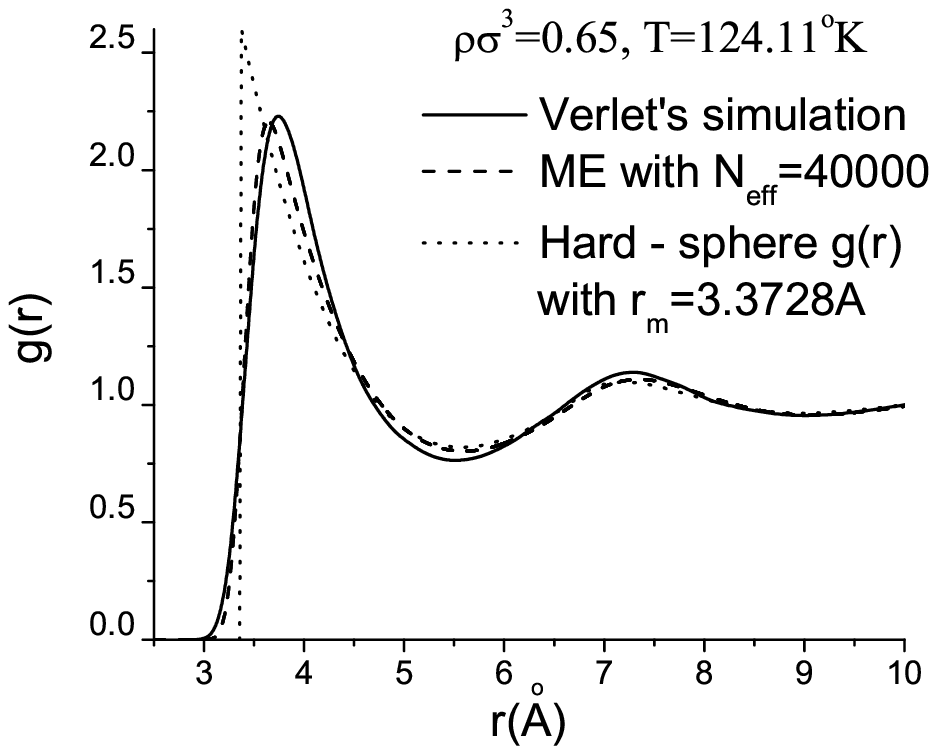}\\
 \begin{picture}(0,155)(0,0)
\put(0,155){(C)}
\end{picture}
 \includegraphics[width=2.0in,height=2.0in]{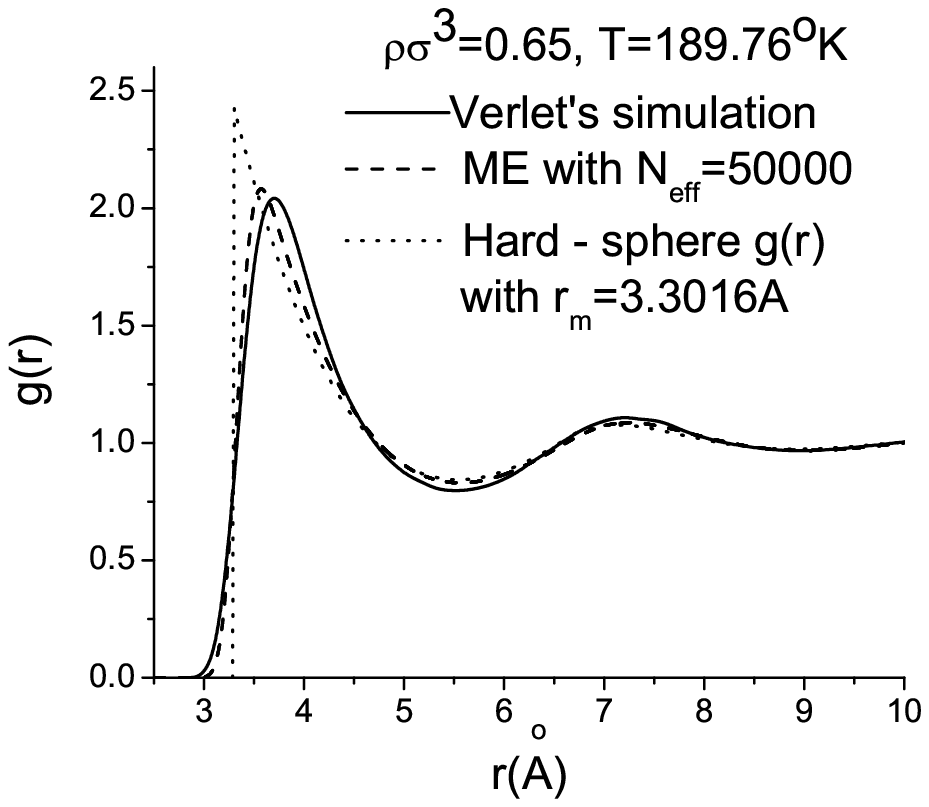}
 \begin{picture}(5,155)(0,0)
\put(0,155){(D)}
\end{picture}
 \includegraphics[width=2.0in,height=2.0in]{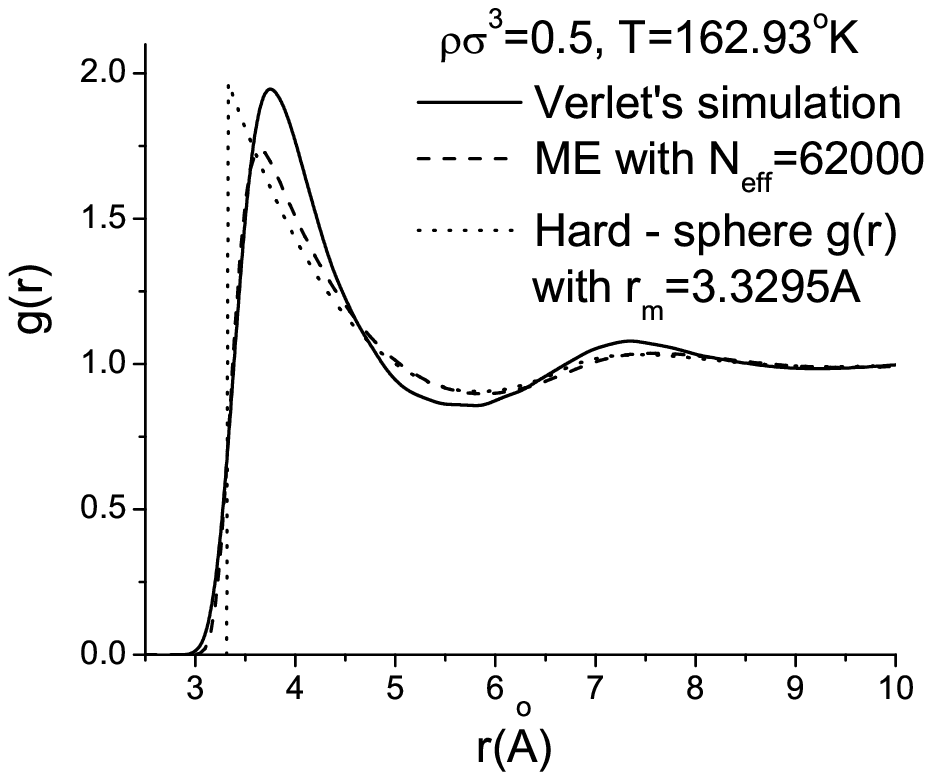}
 \caption{The radial distribution function for (a) the hard-sphere fluid
with optimal diameter $r_{m}$; (b) Verlet's molecular dynamics simulation;
and (c) the improved EME analysis, for argon at (A): density $\protect\rho \protect%
\sigma ^{3}=0.65$, temperature $T=107.82$ $\rm{K}$, and effective particle
number $N_{\rm{eff}}=38000$. (B): $\protect\rho \protect%
\sigma ^{3}=0.65$, $T=124.11$ $\rm{K}$, and $N_{\rm{eff}}=40000$. 
(C): $\protect\rho \protect\sigma ^{3}=0.65$, $T=189.76$ $\rm{K}$, and 
$N_{\rm{eff}}=50000$. (D): $\protect\rho \protect%
\sigma ^{3}=0.5$, $T=162.93$ $\rm{K}$, and $N_{\rm{eff}}=62000$.}
 \label{fig6}
\end{figure}

In figures \ref{fig6}.(A)-(D) we compare three different ways to calculate
the RDF. The solid line is Verlet's molecular dynamics simulation \cite%
{Verlet68}; it plays the role of experimental data against which we compare
our theory. The dotted line is $g_{hs}(r|r_{m})$ for the hard-sphere fluid
with preferred diameter $r_{m}$. This curve, calculated from inverse of %
Eq. (\ref{G(s)}), is also the result of the variational method and coincides with
the ME result for a macroscopically large $N_{\mathrm{eff}}=N$. The dashed
line is the averaged $\bar{g}_{hs}(r)$ of the extended ME analysis. Figures. %
\ref{fig6}.(A)-(C) were plotted at three different temperatures $T=107.82$, $%
124.11$\textrm{\ }and $189.76$ K at the density $\rho \sigma ^{3}=0.65$.
Figure \ref{fig6}.(D) we changed the density and the temperature to $\rho
\sigma ^{3}=0.5$ and $T=162.93$ $K$. The agreement between the ME curve and
Verlet's data is good. The vast improvement over the simpler variational
method calculation is clear. 

One might be tempted to dismiss this achievement as due to the adjustment of
the parameter $N_{\mathrm{eff}}$ but this is not quite correct: $N_{\mathrm{%
eff}}$\ has not been adjusted, it has been calculated by fitting $\bar{g}_{hs}(r)$ to 
optimal mean field RDF $g_{MF}(r)$ for $r<\sigma$. Indeed, despite the fact that the
hard-sphere trial solutions that we employ are mere approximations, the
functional form of the whole curve $\bar{g}_{hs}(r)$ in Eq. (\ref{ghs bar}) is
reproduced quite well. 

However, one may note that the agreement between the ME prediction and Verlet's data becomes worse 
when the fluid density is decreased or the temperature is increased. The reason has been spelled 
out by in the studies of WCA \cite{WCA} They demonstrate that both the 
repulsive and attractive forces contribute to the fluid structure when fluid is at low 
and moderate densities ($ 0.4 \lesssim \rho \sigma^{3} \lesssim 0.65$). 
However, when the fluid density is high 
enough ($\rho \sigma^{3} \gtrsim 0.65$), the repulsive force becomes dominant. 
Because the hard-sphere approximation does not include the attractive force, 
the hard-sphere RDF does not take the attractive force into account, and this error propagates into our
ME prediction. The same discrepancy is also revealed in the WCA theory for
low density \cite{WCA}.

\begin{figure}[ht]
\centering
\includegraphics[width=5.0in,height=3.5in]{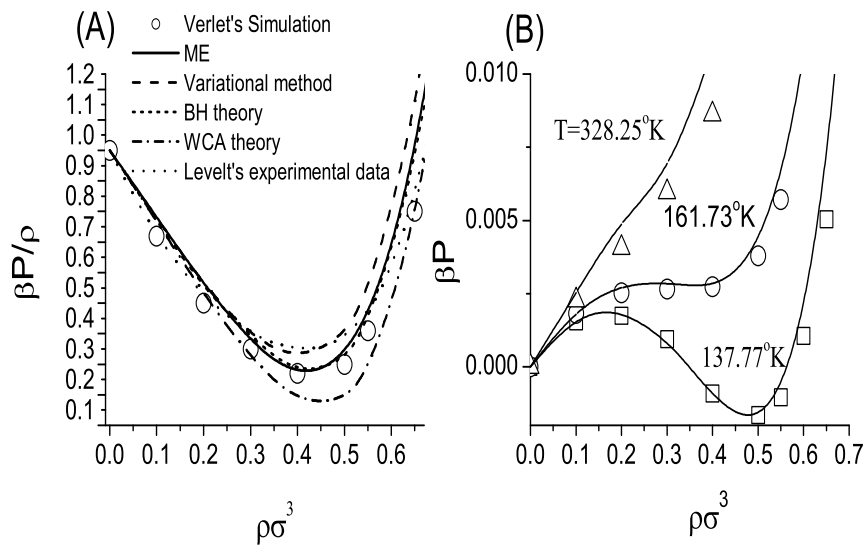}
\caption{(A): The argon equation of state calculated using the EME method, the
variational method and the perturbative theories of BH and WCA are compared
to the Verlet simulation at $T=161.73$ $\rm{K}$. Also shown are Levelt's
experimental results. (B): $\protect\beta P$
versus the reduced density $\protect\rho \protect\sigma ^{3}$ calculated
using the EME method (solid line) and compared to the Hansen-Verlet
simulation for three different isotherms. The graph shows the appearance of
the liquid-gas van der Waals loop as the temperature drops.
}
\label{fig10}
\end{figure}

\noindent \textbf{The equation of state.} 
Finally we use the RDF to calculate the equation of state
from the pressure equation, Eq. (\ref{eq of state}). In Fig \ref{fig10} we compare
the equation of state derived from the $g(r)$ obtained from Verlet's
simulation with calculations using the EME and variational methods and the
perturbative theories of Barker and Henderson \cite{BarkerHenderson76} and
of Weeks, Chandler and Anderson \cite{WCA}, at $T=161.73$ $K$. The EME
results constitute a clear improvement over the plain variational
calculation. For low densities all four methods agree with each other but
differ from the simulation. A better agreement in this region would probably
require a better treatment of two-particle correlations at long distances.
At intermediate densities the best agreement is provided by the EME and BH
results, while the WCA theory seems to be the best at high densities. Also
shown in Fig \ref{fig10}.(A) are experimental data on argon \cite{Levelt60}.
The discrepancy between the experimental curve and the Verlet simulation is
very likely due to the actual potential not being precisely of the
Lennard-Jones type. 

In Fig \ref{fig10}.(B) we plot the EME equation of state for three different
isotherms ($T=137.77$, $161.73$ and $328.25K$). To compare to the simulation
of Hansen and Verlet \cite{HansenVerlet69} we plot $\beta P$ (rather than $%
\beta P/\rho $) as a function of density $\rho \sigma ^{3}$ because this
kind of plot exhibits the characteristic van der Waals loop that signals the
liquid-gas transition as the temperature drops. A more exhaustive
exploration lies, however, outside the scope of this paper.

\section{Conclusion}

\label{sec7} The goal of this paper has been to use the EME method to
generate approximations and show that this provides a generalization of the
Bogoliubov variational principle. This addresses a range of applications
that lie beyond the scope of the traditional MaxEnt. To test the method we
considered simple classical fluids. 

When faced with the difficulty of dealing a system described by an
intractable Hamiltonian, the traditional approach has been to consider a
similar albeit idealized system described by a simpler more tractable
Hamiltonian. The approach we have followed here departs from this tradition:
our goal is not to identify an approximately similar Hamiltonian but rather
to identify an approximately similar probability distribution. The end
result of the EME approach is a probability distribution which is a sum or an
integral over distributions corresponding to different hard-sphere
diameters. While each term in the sum is of a form that can be associated to
a real hard-sphere gas, the sum itself is not of the form $\exp -\beta H$,
and cannot be interpreted as describing any physical system. 

As far as the application to simple fluids is concerned the results achieved
in this paper represent progress but further improvements are possible by
using better approximations to the hard-sphere fluid and by choosing a
broader family of trial distributions. An important improvement would be to
use trial probability distributions that allow inhomogeneities in the hard
sphere diameters. This would lead to a systematic, fully EME method for the
determination of the effective number of particles $N_{\mathrm{eff}}$ that
are locally relevant. 

Many perturbative approaches to fluids had been proposed, and a gradual
process of selection over many years of research led to the optimized
theories of BH and WCA. The variational approach was definitely less
satisfactory than these \textquotedblleft best\textquotedblright\
perturbation theories. With our work, however, the situation has changed:
the EME-improved variational approach offers predictions that already are
competitive with the best perturbative theories. And, of course, the
potential for further improvements of the EME approach remains, at this early
date, far from being exhausted.

\section*{Acknowledgements}
The authors acknowledge R. Scheicher and C.-W. Hong for their valuable
assistance and advice with the numerical calculations.

\newpage

\newpage

\newpage

\newpage

\newpage
\end{document}